# On the reconstruction of block-sparse signals with an optimal number of measurements

*Mihailo Stojnic, Farzad Parvaresh* and *Babak Hassibi*

California Institute of Technology
1200 East California, Pasadena, CA, 91125, USA.
{*mihailo, farzad, hassibi*}@*systems.caltech.edu*

## Abstract

*Let A be an M by N matrix ($M < N$) which is an instance of a real random Gaussian ensemble. In compressed sensing we are interested in finding the sparsest solution to the system of equations $A\mathbf{x} = \mathbf{y}$ for a given $\mathbf{y}$. In general, whenever the sparsity of $\mathbf{x}$ is smaller than half the dimension of $\mathbf{y}$ then with overwhelming probability over A the sparsest solution is unique and can be found by an exhaustive search over $\mathbf{x}$ with an exponential time complexity for any $\mathbf{y}$. The recent work of Candés, Donoho, and Tao shows that minimization of the $\ell_1$ norm of $\mathbf{x}$ subject to $A\mathbf{x} = \mathbf{y}$ results in the sparsest solution provided the sparsity of $\mathbf{x}$, say K, is smaller than a certain threshold for a given number of measurements. Specifically, if the dimension of $\mathbf{y}$ approaches the dimension of $\mathbf{x}$, the sparsity of $\mathbf{x}$ should be $K < 0.239N$. Here, we consider the case where $\mathbf{x}$ is d-block sparse, i.e., $\mathbf{x}$ consists of $n = N/d$ blocks where each block is either a zero vector or a nonzero vector. Instead of $\ell_1$-norm relaxation, we consider the following relaxation*

$$\min_{\mathbf{x}} \|\mathbf{X}_1\|_2 + \|\mathbf{X}_2\|_2 + \ldots + \|\mathbf{X}_n\|_2, \quad \text{subject to} \quad A\mathbf{x} = \mathbf{y} \qquad (\star)$$

*where $\mathbf{X}_i = (\mathbf{x}_{(i-1)d+1}, \mathbf{x}_{(i-1)d+2}, \ldots, \mathbf{x}_{id})$ for $i = 1, 2, \ldots, N$. Our main result is that as $n \to \infty$, ($\star$) finds the sparsest solution to $A\mathbf{x} = \mathbf{y}$, with overwhelming probability in A, for any $\mathbf{x}$ whose block sparsity is $k/n < \frac{1}{2} - O(\varepsilon)$, provided $M/N > 1 - 1/d$, and $d = \Omega(\log(1/\varepsilon)/\varepsilon)$. The relaxation given in ($\star$) can be solved in polynomial time using semi-definite programming.*

## 1. Introduction

Let $A$ be an $M$ by $N$ instance of the real random Gaussian ensemble and $\mathbf{x}$ be an $N$ dimensional signal from $\mathbb{R}^N$ with sparsity $K$, i.e., only $K$ elements of $\mathbf{x}$ are nonzero. Set $\mathbf{y} = A\mathbf{x}$ which is an $M$ dimensional vector in $\mathbb{R}^M$. In compressed sensing $\mathbf{y}$ is called the *measurement vector* and $A$ the *Gaussian measurement* matrix. Compressed sensing has applications in many different fields such as data mining [14], error-correcting codes [12, 16, 18], DNA microarrays [13, 33, 34], astronomy, tomography, digital photography, and A/D converters.

In general, when $K \ll N$ one can hope that $\mathbf{y} = A\mathbf{x}$ is unique for large enough $M$ which is much smaller than $N$. In other words, instead of sensing an $N$ dimensional signal $\mathbf{x}$ with sparsity $K$ we can measure $M$ random linear functionals of $\mathbf{x}$ where $M \ll N$ and find $\mathbf{x}$ by solving the under-determined system of equations $\mathbf{y} = A\mathbf{x}$ with the extra condition that $\mathbf{x}$ is $K$ sparse. The reconstruction can be presented as the following optimization problem:

$$\min_{\mathbf{x}} \|\mathbf{x}\|_0 \quad \text{subject to } A\mathbf{x} = \mathbf{y} \qquad (1)$$

where the $\ell_0$ norm or the Hamming norm is the number of nonzero elements of **x**.

Define $\alpha \stackrel{\text{def}}{=} M/N$ and $\beta \stackrel{\text{def}}{=} K/N$. In [15], the authors show that if $\beta > \frac{1}{2}\alpha$ then for any measurement matrix $A$ one can construct different $K$ sparse signals $\mathbf{x}_1$ and $\mathbf{x}_2$ such that $A\mathbf{x}_1 = A\mathbf{x}_2$. In addition, if $\beta \leqslant \frac{1}{2}\alpha$ then there exists an $A$ such that the $K$ sparse solution to $\mathbf{y} = A\mathbf{x}$ is unique for any $\mathbf{y}$; specifically, for random Gaussian measurements the uniqueness property holds with overwhelming probability in the choice of $A$. However, the reconstruction of **x** for a given **y** can be cumbersome. One of the fundamental questions in compressed sensing is whether one can *efficiently* recover **x** using an optimal number of measurements.

## 1.1. Prior work

A naive exhaustive search can reconstruct the $K$ sparse solution **x** to the systems of equations $\mathbf{y} = A\mathbf{x}$ with $O\left(\binom{N}{K}M^3\right)$ complexity. Recently, Candés, Romberg, Tao and Donoho [10, 11, 30], show that the $\ell_0$ optimization can be relaxed to $\ell_1$ minimization if the sparsity of the signal is $K = O(M/\log(N/M))$. In this case, the sparse signal is the solution to the following $\ell_1$ norm optimization with high probability in the choice of $A$:

$$\min_{\mathbf{x}} \|\mathbf{x}\|_1 \quad \text{subject to} \quad A\mathbf{x} = \mathbf{y} \tag{2}$$

This optimization can be solved efficiently using linear programming. Faster algorithms were discovered in [1–3, 35]. For a comprehensive list of papers and results in compressed sensing please check [4].

Donoho and Tanner [5, 7, 8] determined the region $(\alpha, \beta)$ for which the $\ell_1$ and $\ell_0$ coincide under Gaussian measurements for every (or almost every) $K$-sparse vector **x**. From a refinement of their result given in [45], when $\beta$ approaches $\alpha$ the sparsity has to be smaller than $0.239N$. Notice that, ideally, one should be able to recover the signal if the sparsity is less than $\frac{1}{2}N$. We have to mention that with Vandermonde measurements we can recover the sparse signal with optimal number of measurements efficiently [15]. However, it is not clear whether the resulting algorithms (which are variations of recovering a measure from its moments) are robust with respect to imperfections, such as noise [9, 27–29]. Also, results similar to those valid for Gaussian matrices $A$ have been established for several different ensembles, e.g., Fourier (see e.g., [11]).

In this paper, we will focus on developing robust efficient algorithms that work for Gaussian measurements.

## 1.2. Our main result

We consider the reconstruction of **block-sparse** signals from their random measurements. A signal of dimension $N$ which consists of $n$ blocks of size $d = N/n$ is $k$ sparse if only $k$ blocks of the signal out of $n$ are nonzero. Such signals arise in various applications, e.g., DNA microarrays, equalization of sparse communication channels, magnetoencephalography etc. (see e.g., [33, 34, 36, 39–41] and the references therein). We measure the signal with a $md \times nd$ random Gaussian matrix $\mathbf{y} = A\mathbf{x}$. More on a scenario similar to this one the interested reader can find in e.g. [36–38, 42, 44]. Using the $\ell_1$ relaxation for reconstructing **x** does not exploit the fact that the signal is block-sparse, i.e. that the nonzero entries occur in consecutive positions. Instead, different techniques were used throughout the literature. In [36] the authors adapt standard orthogonal matching pursuit algorithm (used normally in case $k = 1$) to the block-sparse case. In [37, 38, 42, 43] the authors use certain convex or non-convex relaxations (mostly different from the standard $\ell_1$) and discuss their performances. Generalization of the block-sparse problem to the case when the number of blocks is infinite was considered in the most recent paper [44]. In this paper we consider the following convex relaxation for the recovery of **x**:

$$\min_{\mathbf{x}} \|\mathbf{X}_1\|_2 + \|\mathbf{X}_2\|_2 + \cdots + \|\mathbf{X}_n\|_2, \quad \text{subject to} \quad A\mathbf{x} = \mathbf{y} \tag{3}$$

where $\mathbf{X}_i = (\mathbf{x}_{(i-1)d+1}, \mathbf{x}_{(i-1)d+2}, \ldots, \mathbf{x}_{id})$, for $i = 1, 2, \ldots, n$. We will analyze its theoretical performance and show that for a large enough $d$, independent of $n$, if $\alpha$ approaches one, $\beta$ can approach $\frac{1}{2}$ and the optimization of

(3) will give the unique sparse solution with overwhelming probability over the choice of $A$ for any $\mathbf{y}$. We will also briefly outline how (3) can be posed as a semi-definite program and therefore solved efficiently in polynomial time by a host of numerical methods. Furthermore, we will demonstrate how (3) can be adapted for practical considerations. Numerical results that we will present indicate that in practice a modified version of (3) (given in Section 4) will even for moderate values of $d$ be able to recover most of the signals with sparsity fairly close to the number of measurements. Before proceeding further we state the main result of this work in the following theorem.

**Theorem 1.** *Let $A$ be an $md \times nd$ matrix. Further, let $A$ be an instance of the random Gaussian ensemble. Assume that $\epsilon$ is a small positive number, i.e., $0 < \epsilon \ll 1$, $d = \Omega(\log(1/\epsilon)/\epsilon)$, $\alpha > 1 - 1/d$, and $\beta = \frac{1}{2} - O(\epsilon)$. Also, assume that $n$ tends to infinity, $m = \alpha n$, and the block-sparsity of $\mathbf{x}$ is smaller than $\beta n$. Then, with overwhelming probability, any $d$-block sparse signal $\mathbf{x}$ can be reconstructed efficiently from $\mathbf{y} = A\mathbf{x}$ by solving the optimization problem (3).*

Our proof technique does not use the restricted isometry property of the measurement matrix $A$, introduced in the work of Candés and Tao [11] and further discussed in [17], nor does it rely on the $k$-neighborliness of the projected polytopes presented in the work of Donoho and Tanner [5, 7, 8, 19]. Instead, we look at the null-space of the measurement matrix $A$ and use a generalization of a necessary and sufficient condition given in [31] for the equivalence of (1) and (3).

We are able to use some probabilistic arguments to show that, for a random Gaussian measurement matrix, (4) given below holds with overwhelming probability. In our proof we use a union bound to upper bound the probability that (4) fails; this makes our bound loose for $\alpha$ less than one. We expect to get sharp thresholds for other values of $\alpha$ by generalizing the idea of looking at the neighborliness of randomly projected simplices presented in [5, 7, 8]. However, for relaxation in (3) instead of simplices we have to work with the convex hull $\mathcal{B}$ of $n$ $d$-dimensional spheres. Specifically, one would need to compute the probability that a random $h$-dimensional affine plane that passes through a point on the boundary of $\mathcal{B}$ will be inside the tangent cone of that given point. Solving this problem seems to be rather difficult.

## 2. Null-space characterization

In this section we introduce a necessary and sufficient condition on the measurement matrix $A$ so that the optimizations of (1) and (3) are equivalent for all $k$-block sparse $\mathbf{x}$. (see [24–26, 31] for variations of this result). Throughout the paper we set $\mathcal{H}$ to be the set of all subsets of size $k$ of $\{1, 2, \ldots, n\}$ and by $\bar{\mathcal{K}}$ we mean the complement of the set $\mathcal{K} \subset \mathcal{H}$ with respect to $\{1, 2, \ldots, n\}$, i.e., $\bar{\mathcal{K}} = \{1, 2, \ldots, n\} \setminus \mathcal{K}$.

**Theorem 2.** *Assume that $A$ is a $dm \times dn$ measurement matrix, $\mathbf{y} = A\mathbf{x}$ and $\mathbf{x}$ is $k$-block sparse. Then (3) coincides with the solution of (1) if and only if for all nonzero $\mathbf{w} \in \mathbb{R}^{dn}$ where $A\mathbf{w} = 0$ and all $\mathcal{K} \in \mathcal{H}$*

$$\sum_{i \in \mathcal{K}} ||\mathbf{W}_i||_2 < \sum_{i \in \bar{\mathcal{K}}} ||\mathbf{W}_i||_2 \tag{4}$$

*where $\mathbf{W}_i = (\mathbf{w}_{(i-1)d+1}, \mathbf{w}_{(i-1)d+2}, \ldots, \mathbf{w}_{id})$, for $i = 1, 2, \ldots, n$.*

*Proof.* The proof goes along the same line as the proofs in [24–26, 31]. The only difference is that each component of the vector is now replaced by the two norm of the subvector. First we prove that if (4) is satisfied then the solution of (3) coincides with the solution (1). Let $\bar{\mathbf{x}}$ be the solution of (1) and let $\hat{\mathbf{x}}$ be the solution of (3). Further, let $\bar{\mathbf{X}}_i = (\bar{\mathbf{x}}_{(i-1)d+1}, \bar{\mathbf{x}}_{(i-1)d+2}, \ldots, \bar{\mathbf{x}}_{id})$, for $i = 1, 2, \ldots, n$ and $\hat{\mathbf{X}}_i = (\hat{\mathbf{x}}_{(i-1)d+1}, \hat{\mathbf{x}}_{(i-1)d+2}, \ldots, \hat{\mathbf{x}}_{id})$, for

$i = 1, 2, \ldots, n$. Set $\mathcal{K}$ to be the support of $\bar{\mathbf{x}}$, then we can write

$$\begin{aligned}
\sum_{i=1}^{n} ||\hat{\mathbf{X}}_i||_2 &= \sum_{i=1}^{n} ||\hat{\mathbf{X}}_i - \bar{\mathbf{X}}_i + \bar{\mathbf{X}}_i||_2 \\
&= \sum_{i \in \mathcal{K}} ||\hat{\mathbf{X}}_i - \bar{\mathbf{X}}_i + \bar{\mathbf{X}}_i||_2 + \sum_{i \in \bar{\mathcal{K}}} ||\hat{\mathbf{X}}_i - \bar{\mathbf{X}}_i + \bar{\mathbf{X}}_i||_2 \\
&= \sum_{i \in \mathcal{K}} ||\hat{\mathbf{X}}_i - \bar{\mathbf{X}}_i + \bar{\mathbf{X}}_i||_2 + \sum_{i \in \bar{\mathcal{K}}} ||\hat{\mathbf{X}}_i - \bar{\mathbf{X}}_i||_2 \\
&\geqslant \sum_{i=1}^{n} ||\bar{\mathbf{X}}_i||_2 - \sum_{i \in \mathcal{K}} ||\hat{\mathbf{X}}_i - \bar{\mathbf{X}}_i||_2 + \sum_{i \in \bar{\mathcal{K}}} ||\hat{\mathbf{X}}_i - \bar{\mathbf{X}}_i||_2. \quad (5)
\end{aligned}$$

Since $\bar{\mathbf{x}} - \hat{\mathbf{x}}$ lies in the null-space of $A$, we have $\sum_{i \in \mathcal{K}} ||\hat{\mathbf{X}}_i - \bar{\mathbf{X}}_i||_2 < \sum_{i \in \bar{\mathcal{K}}} ||\hat{\mathbf{X}}_i - \bar{\mathbf{X}}_i||_2$. Thus, (5) implies $\sum_{i=1}^{n} ||\hat{\mathbf{X}}_i||_2 > \sum_{i=1}^{n} ||\bar{\mathbf{X}}_i||_2$, which is a contradiction. Therefore, $\bar{\mathbf{x}} = \hat{\mathbf{x}}$. Now we prove the converse. Assume (4) does not hold. Then there exists $\mathbf{w} \in \mathbb{R}^{nd}$, $A\mathbf{w} = 0$, $\mathbf{w} = \binom{\mathbf{w}_1}{\mathbf{w}_2}$, $\mathbf{w}_1 \in \mathbb{R}^{kd}$, $\mathbf{w}_2 \in \mathbb{R}^{(n-k)d}$ such that $\mathbf{w}_1$ is $k$-block sparse and $\sum_{i \in \mathcal{K}} ||\mathbf{W}_i||_2 \leqslant \sum_{i \in \bar{\mathcal{K}}} ||\mathbf{W}_i||_2$, where $\mathcal{K}$ is the support of $\mathbf{w}_1$. Take $\mathbf{x} = \binom{\mathbf{w}_1}{\mathbf{0}}$ and $\mathbf{y} = A\mathbf{x}$. Since $\mathbf{w}$ is in the null-space of $A$, $\mathbf{y} = A\binom{\mathbf{0}}{-\mathbf{w}_2}$. Therefore we have found a signal $\binom{\mathbf{0}}{-\mathbf{w}_2}$ which is not $k$-block sparse and has smaller norm than the $k$-block sparse $\binom{\mathbf{w}_1}{\mathbf{0}}$. ∎

**Remark.** We need not to check (4) for all subsets $\mathcal{K}$; checking the subset with the $k$ largest (in two norm) blocks of $\mathbf{w}$ is sufficient. However, the form of Theorem 2 will be more convenient for our subsequent analysis.

Let $Z$ be a basis of the null space of $A$, so that any $dn$ dimensional vector $\mathbf{w}$ in the the null-space of $A$ can be represented as $Z\mathbf{v}$ where $\mathbf{v} \in \mathbb{R}^{d(n-m)}$. For any $\mathbf{v} \in \mathbb{R}^{d(n-m)}$ write $\mathbf{w} = Z\mathbf{v}$. We split $\mathbf{w}$ into blocks of size $d$, $\mathbf{W}_i = (\mathbf{w}_{(i-1)d+1}, \mathbf{w}_{(i-1)d+2}, \ldots, \mathbf{w}_{id})$, for $i = 1, 2, \ldots, n$. Then, the condition (4) of Theorem 2 is equivalent to

$$\sum_{i \in \mathcal{K}} \mathbf{W}_i \leqslant \sum_{i \in \bar{\mathcal{K}}} \mathbf{W}_i, \text{ for any } \mathbf{v} \in \mathbb{R}^{d(n-m)} \text{ and } \mathcal{K} \in \mathscr{K}, \text{ where } \mathbf{w} = Z\mathbf{v}. \quad (6)$$

We denote by $I_\mathbf{v}$ the event that (6) happens. In the following we find an upper bound on the probability that $I_\mathbf{v}$ fails as $n$ tends to infinity. We will show that for certain values of $\alpha$, $\beta$, and $d$ this probability tends to zero.

**Lemma 3.** *Let $A \in R^{dm \times dn}$ be a random matrix with i.i.d. $\mathcal{N}(0, 1)$ entries. Then the following statements hold:*

- *The distribution of $A$ is left-rotationally invariant, $P_A(A) = P_A(A\Theta)$, $\Theta\Theta^* = \Theta^*\Theta = I$*

- *The distribution of $Z$, any basis of the null-space of $A$ is right-rotationally invariant. $P_Z(Z) = P_Z(\Theta^* Z)$, $\Theta\Theta^* = \Theta^*\Theta = I$*

- *It is always possible to choose a basis for the null-space such that $Z \in R^{dn \times d(n-m)}$ has i.i.d. $\mathcal{N}(0, 1)$ entries.*

In view of Theorem 2 and Lemma 3, for any $A$ whose null-space is rotationally invariant the sharp bounds of [6], for example, apply (of course, if $k = 1$). In this paper, we shall analyze the null-space directly.

## 3. Probabilistic analysis of the null-space characterization

Assume $Z$ is an $dn \times d(n - m)$ matrix whose components are i.i.d. zero-mean unit-variance Gaussian random variables. Define $Z_i$ to be the matrix which consists of the $\{(i-1)d + 1, (i-1)d + 2, \ldots, id\}$ rows of $Z$ and define $Z_{ij}$ to be the $j$-th column of $Z_i$. Let $\alpha = 1 - \gamma, 0 < \gamma \ll 1$ where $\gamma$ is a constant independent of $n$. Then we will find a $d$ such that $\beta \to \frac{1}{2}$ and
$$\lim_{n \to \infty} P(I_{\mathbf{v}}) = 1. \tag{7}$$
Proving (7) is enough to show that for all random matrix ensembles which have isotropically distributed nullspace, (3) with overwhelming probability solves (1). In order to prove (7) we will actually look at the complement of the event $I_{\mathbf{v}}$ and we show that
$$\lim_{n \to \infty} P_f \stackrel{\text{def}}{=} \lim_{n \to \infty} P(\bar{I}_{\mathbf{v}}) = 0, \tag{8}$$
where $\bar{I}_{\mathbf{v}}$ denotes the complement of the event $I_{\mathbf{v}}$. Using the union bound we can write
$$P_f \leqslant \sum_{\mathcal{K} \in \mathscr{K}} P\left(\exists \mathbf{v} \in \mathbb{R}^{d(n-m)} : \sum_{i \in \mathcal{K}} ||Z_i \mathbf{v}||_2 \geqslant \sum_{i \in \bar{\mathcal{K}}} ||Z_i \mathbf{v}||_2\right) \tag{9}$$
Clearly the size of $\mathscr{K}$ is $\binom{n}{k}$. Since the probability in (9) is insensitive to scaling of $\mathbf{v}$ by a constant we can restrict $\mathbf{v}$ to lie on the surface of a shape $\mathcal{C}$ that encapsulates the origin. Furthermore, since the elements of the matrix $Z$ are i.i.d. all $\binom{n}{k}$ terms in the first summation on the right hand side of (9) will then be equal. Therefore we can further write
$$P_f \leqslant \binom{n}{k} \cdot P\left(\exists \mathbf{v} \in \mathcal{C} : \sum_{i=1}^{k} ||Z_i \mathbf{v}||_2 \geqslant \sum_{i=k+1}^{n} ||Z_i \mathbf{v}||_2\right). \tag{10}$$
The main difficulty in computing the probability on the right hand side of (10) is in the fact that the vector $\mathbf{v}$ is continuous. Our approach will be based on the discrete covering of the unit sphere. In order to do that we will use small spheres of radius $\epsilon$. It can be shown [18, 20, 21] that $\epsilon^{-d(n-m)}$ spheres of radius $\epsilon$ is enough to cover the surface of the $d(n-m)$-dimensional unit sphere. Let the coordinates of the centers of these $\epsilon^{-d(n-m)}$ small spheres be the vectors $\mathbf{z}_t, t = 1, 2, \ldots, \epsilon^{-d(n-m)}$. Clearly, $||\mathbf{z}_t||_2 = \sqrt{1-\epsilon^2}$. Further, let $S_t, t = 1, 2, \ldots, \epsilon^{-d(n-m)}$ be the intersection of the unit sphere and the hyperplane through $\mathbf{z}_t$ perpendicular on the line that connects $\mathbf{z}_t$ and the origin. It is not difficult to see that $\bigcup_{t=1}^{\epsilon^{-d(n-m)}} S_t$ forms a body which completely encapsulates the origin. This effectively means that for any point $\mathbf{v}$ such that $||\mathbf{v}|| > 1$, the line connecting $\mathbf{v}$ and the origin will intersect $\bigcup_{t=1}^{\epsilon^{-d(n-m)}} S_t$. Hence, we set $\mathcal{C} = \bigcup_{t=1}^{\epsilon^{-d(n-m)}} S_t$ and apply union bound over $S_t$ to get
$$P_f \leqslant \binom{n}{k} \epsilon^{-d(n-m)} \max_t \left[ P\left(\exists \mathbf{v} \in S_t : \sum_{i=1}^{k} ||Z_i \mathbf{v}||_2 \geqslant \sum_{i=k+1}^{n} ||Z_i \mathbf{v}||_2\right) \right]. \tag{11}$$
Every vector $\mathbf{v} \in S_t$ can be represented as $\mathbf{v} = \mathbf{z}_t + \mathbf{e}$ where $||\mathbf{e}||_2 \leqslant \epsilon$. Then we have
$$\max_t \left[ P\left(\exists \mathbf{v} \in S_t : \sum_{i=1}^{k} ||Z_i \mathbf{v}||_2 \geqslant \sum_{i=k+1}^{n} ||Z_i \mathbf{v}||_2\right) \right]$$
$$= \max_t \left[ P\left(\exists \mathbf{e} : ||\mathbf{e}||_2 \leqslant \epsilon \text{ and } \sum_{i=1}^{k} ||Z_i(\mathbf{z}_t + \mathbf{e})||_2 \geqslant \sum_{i=k+1}^{n} ||Z_i(\mathbf{z}_t + \mathbf{e})||_2\right) \right]. \tag{12}$$
Given the symmetry of the problem (i.e. the rotaional invariance of the $Z_i$) it should be noted that, without loss of generality, we can assume $\mathbf{z}_t = [||\mathbf{z}_t||_2, 0, 0, \ldots, 0]$. Further, using the results from [23] we have that $\eta^{d(n-m)-1}$ points can be located on the sphere of radius $c\epsilon$ centered at $\mathbf{z}_t$ such that $S_t$ (which lies in a $(d(n-m)-1)$-dimensional space and whose radius is $\epsilon$) is inside a polytope determined by them and
$$c \leqslant \begin{cases} \frac{1}{(1-\ln(\eta))\sqrt{2\ln(\eta) - \frac{\ln(d(n-m)-1)}{d(n-m)-1}}} & \text{if } \eta < \sqrt{2} \\ \frac{1}{1-(1+\frac{1}{\eta^2})\frac{1}{2\eta^2}} & \text{otherwise.} \end{cases} \tag{13}$$

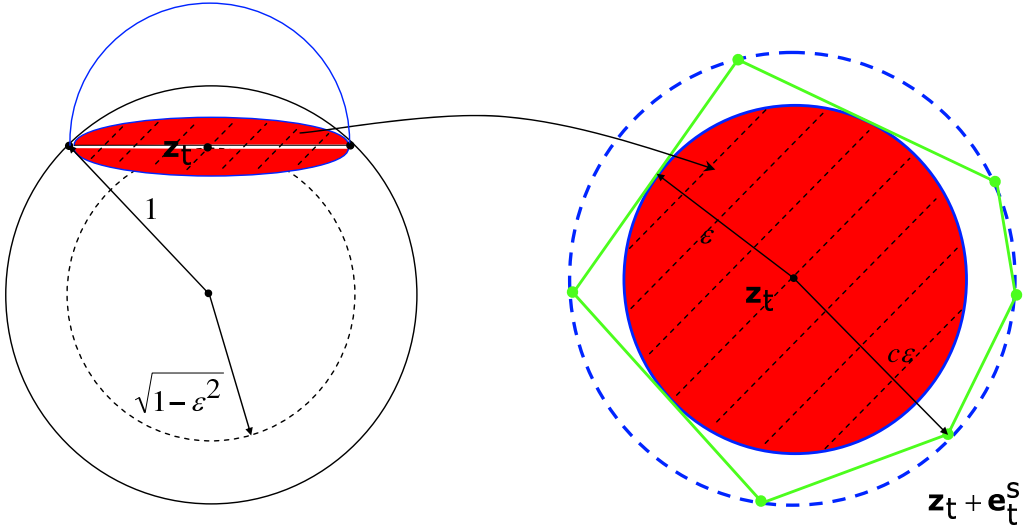

**Figure 1. Covering of the unit sphere**

To get a feeling for what values $\eta$ and $c$ can take we refer to [22] where it was stated that $3^{d(n-m)-1}$ points can be located on the sphere of radius $\sqrt{\frac{9}{8}}\epsilon$ centered at $\mathbf{z}_t$ such that $S_t$ is inside a polytope determined by them.

Let us call the polytope determined by $\eta^{d(n-m)-1}$ points $P_t$. Let $\mathbf{e}_t^s, s = 1, 2, \ldots, \eta^{d(n-m)-1}$ be its $\eta^{d(n-m)-1}$ corner points. Since $||Z_i \mathbf{z}_t||_2 - ||Z_i \mathbf{e}||_2 \leq ||Z_i(\mathbf{z}_t + \mathbf{e})||_2$, and $S_t \subset P_t$ we have

$$\max_t P(\exists \mathbf{e}, ||\mathbf{e}||_2 \leq \epsilon \text{ s. t. } \sum_{i=1}^{k} ||Z_i(\mathbf{z}_t + \mathbf{e})||_2 \geq \sum_{i=k+1}^{n} ||Z_i(\mathbf{z}_t + \mathbf{e})||_2)$$

$$\leq \max_t P(\exists \mathbf{e}, (\mathbf{z}_t + \mathbf{e}) \in P_t \text{ s. t. } \sum_{i=1}^{k} ||Z_i(\mathbf{z}_t + \mathbf{e})||_2 \geq \sum_{i=k+1}^{n} (||Z_i \mathbf{z}_t||_2 - ||Z_i \mathbf{e}||_2))$$

$$\leq \max_t P\left(\max_s \left(\sum_{i=k+1}^{n} ||Z_i \mathbf{e}_t^s||_2 + \sum_{i=1}^{k} ||Z_i(\mathbf{z}_t + \mathbf{e}_t^s)||_2\right) \geq \sum_{i=k+1}^{n} ||Z_i \mathbf{z}_t||_2\right). \tag{14}$$

where the second inequality follows from the property that the maximum of a convex function over a polytope is achieved at its corner points and that function inside the $\max_s$ is convex as it is a sum of convex norms. Connecting (11), (12), and (14) we obtain

$$P_f \leq \frac{\binom{n}{k}}{\epsilon^{d(n-m)}} \max_t P\left(\max_s \left(\sum_{i=k+1}^{n} ||Z_i \mathbf{e}_t^s||_2 + \sum_{i=1}^{k} ||Z_i(\mathbf{z}_t + \mathbf{e}_t^s)||_2\right) \geq \sum_{i=k+1}^{n} ||Z_i \mathbf{z}_t||_2\right). \tag{15}$$

Using the union bound over $s$ we further have

$$\max_t P\left(\max_s \left(\sum_{i=k+1}^{n} ||Z_i \mathbf{e}_t^s||_2 + \sum_{i=1}^{k} ||Z_i(\mathbf{z}_t + \mathbf{e}_t^s)||_2\right) \geq \sum_{i=k+1}^{n} ||Z_i \mathbf{z}_t||_2\right)$$

$$\leq \max_t \sum_{s'=1}^{\eta^{d(n-m)-1}} P\left(\left(\sum_{i=k+1}^{n} ||Z_i \mathbf{e}_t^{s'}||_2 + \sum_{i=1}^{k} ||Z_i(\mathbf{z}_t + \mathbf{e}_t^{s'})||_2\right) \geq \sum_{i=k+1}^{n} ||Z_i \mathbf{z}_t||_2\right). \tag{16}$$

Given that only the first component of $\mathbf{z}_t$ is not equal to zero and the symmetry of the problem we can write

$$\max_t \eta^{d(n-m)-1} \sum_{s'=1} P\left(\left(\sum_{i=k+1}^n ||Z_i \mathbf{e}_t^{s'}||_2 + \sum_{i=1}^k ||Z_i(\mathbf{z}_t + \mathbf{e}_t^{s'})||_2\right) \geqslant \sum_{i=k+1}^n ||Z_i \mathbf{z}_t||_2\right)$$

$$\leqslant \eta^{d(n-m)-1} \max_{t,s'} P\left(\left(\sum_{i=k+1}^n ||\sum_{j=2}^{d(n-m)} Z_{ij}(\mathbf{e}_t^{s'})_j||_2 + \sum_{i=1}^k ||Z_i(\mathbf{z}_t + \mathbf{e}_t^{s'})||_2\right) \geqslant \sum_{i=k+1}^n ||Z_i||_2(||\mathbf{z}_t||_2 - |(\mathbf{e}_t^{s'})_1|)\right) \quad (17)$$

where $(\mathbf{e}_t^{s'})_j$ denotes $j$-th components of $\mathbf{e}_t^{s'}$. Let $B_i = Z_i(\mathbf{z}_t + e_t^{s'})$, $C_i = Z_{i1}(||\mathbf{z}_t||_2 - |(\mathbf{e}_t^{s'})_1|)$, and $D_i = \sum_{j=2}^{d(n-m)} Z_{ij}(\mathbf{e}_t^{s'})_j$. Clearly, $B_i$, $C_i$, $D_i$ are independent zero-mean Gaussian random vectors of length $d$. Then we can rewrite (17) as

$$\max_t \eta^{d(n-m)-1} \sum_{s'=1} P\left(\left(\sum_{i=k+1}^n ||Z_i \mathbf{e}_t^{s'}||_2 + \sum_{i=1}^k ||Z_i(\mathbf{z}_t + \mathbf{e}_t^{s'})||_2\right) \geqslant \sum_{i=k+1}^n ||Z_i \mathbf{z}_t||_2\right)$$

$$\leqslant (\eta)^{d(n-m)-1} \max_{t,s'} P\left(\sum_{i=k+1}^n ||D_i||_2 + \sum_{i=1}^k ||B_i||_2 \geqslant \sum_{i=k+1}^n ||C_i||_2\right). \quad (18)$$

Let $B_{i_p}$, $C_{i_p}$, and $D_{i_p}$ denote the $p$-th components of the vectors $B_i$, $C_i$, $D_i$, respectively. Then for any $1 \leqslant p \leqslant d$ it holds

$$\operatorname{var}(B_{i_p}) = ||\mathbf{z}_t + \mathbf{e}_t^{s'}||_2^2 = 1 - \epsilon^2 + c\epsilon^2, \operatorname{var}(C_{i_p}) = (||\mathbf{z}_t||_2 - |(\mathbf{e}_t^{s'})_1|)^2, \operatorname{var}(D_{i_p}) = ||\mathbf{e}_t^{s'}||_2^2 - |(\mathbf{e}_t^{s'})_1|^2.$$

Let $G_i$, $F_i$ be independent zero-mean Gaussian random vectors such that such that for any $1 \leqslant p \leqslant d$

$$\operatorname{var}(G_{i_p}) = (||\mathbf{z}||_2 - ||\mathbf{e}_t^{s'}||_2)^2, \operatorname{var}(F_{i_p}) = ||\mathbf{e}_t^{s'}||_2^2.$$

Since $\operatorname{var}(G_{i_p}) \leqslant \operatorname{var}(C_{i_p})$, and $\operatorname{var}(F_{i_p}) \geqslant \operatorname{var}(D_{i_p})$ we have from (18)

$$\eta^{d(n-m)-1} \max_{t,s'} P\left(\sum_{i=k+1}^n ||D_i||_2 + \sum_{i=1}^k ||B_i||_2 \geqslant \sum_{i=k+1}^n ||C_i||_2\right)$$

$$\leqslant \eta^{d(n-m)-1} \max_{t,s'} P\left(\sum_{i=k+1}^n ||F_i||_2 + \sum_{i=1}^k ||B_i||_2 \geqslant \sum_{i=k+1}^n ||G_i||_2\right). \quad (19)$$

Since $||\mathbf{e}_t^{s'}||_2$ does not depend on $t$, $s'$, the outer maximization can be omitted. Furthermore, $||\mathbf{e}_t^{s'}||_2 = c\epsilon$. Using the Chernoff bound we further have

$$\eta^{d(n-m)-1} P\left(\sum_{i=1}^k ||B_i||_2 \geqslant \sum_{i=k+1}^n (||G_i||_2 - ||F_i||_2)\right)$$

$$\leqslant \eta^{d(n-m)-1} (Ee^{\mu ||B_1||_2})^k (Ee^{-\mu ||G_1||_2})^{n-k} (Ee^{\mu ||F_1||_2})^{n-k}. \quad (20)$$

where $\mu$ is a positive constant. Connecting (15)-(20) we have

$$P_f \leqslant \binom{n}{k} \frac{1}{\eta} \left(\frac{\eta}{\epsilon}\right)^{d(n-m)} (Ee^{\mu ||B_1||_2})^k \left(\frac{Ee^{-\mu ||G_1||_2}}{(Ee^{\mu ||F_1||_2})^{-1}}\right)^{n-k}. \quad (21)$$

After setting $k = \beta n$, $m = \alpha n$, and using the fact that $\binom{n}{k} \approx e^{-nH(\beta)}$ we finally obtain

$$\lim_{n\to\infty} P_f \leq \lim_{n\to\infty} \xi^n \tag{22}$$

where

$$\xi = \frac{(\eta/\epsilon)^{d(1-\alpha)}}{e^{H(\beta)}} (Ee^{\mu\|B_1\|_2})^\beta \left(\frac{Ee^{-\mu\|G_1\|_2}}{(Ee^{\mu\|F_1\|_2})^{-1}}\right)^{1-\beta}. \tag{23}$$

and $H(\beta) = \beta \ln \beta + (1-\beta)\ln(1-\beta)$. We now set $\mu = \sqrt{2d-1}\delta\sqrt{2}, \delta \ll 1$. In the appendices we will determine $Ee^{\sqrt{2d-1}\delta\sqrt{2}\|B_1\|_2}$, $Ee^{\sqrt{2d-1}\delta\sqrt{2}\|F_1\|_2}$, and $Ee^{-\sqrt{2d-1}\delta\sqrt{2}\|G_1\|_2}$.

We now return to the analysis of (23). Replacing the results from (37), (38), and (44) in (23) we finally have

$$\xi \approx \frac{(\eta/\epsilon)^{d(1-\alpha)}}{e^{H(\beta)}} \left(e^{d((\delta b)^2 + \delta b)}\right)^\beta \left(e^{d((\delta f)^2 + \delta f)}\right)^{1-\beta} \left(e^{d((\delta g)^2 - \delta g)}\right)^{1-\beta} \tag{24}$$

where we recall that $b = \sqrt{1-\epsilon^2+c^2\epsilon^2}$, $f = c\epsilon$, and $g = \sqrt{1-\epsilon^2} - c\epsilon$. Our goal is to find $d$ such that for $\alpha = 1-\gamma, 0 < \gamma \ll 1$ and $\beta = \frac{1}{2} - \sigma, 0 < \sigma \ll \frac{1}{2}, \xi < 1$. That means we need

$$\ln(\xi) < 0 \tag{25}$$

which implies

$$d(1-\alpha)\ln(\frac{\eta}{\epsilon}) + d\delta(\beta b + (1-\beta)f - (1-\beta)g) + d\delta^2(\beta b^2 + (1-\beta)f^2 + (1-\beta)g^2) < H(\beta). \tag{26}$$

Let

$$\beta_{\text{opt}} = \frac{g-f}{g+b-f} \approx \frac{1-2c\epsilon}{2-2c\epsilon} \tag{27}$$

Combining the previous results the following theorem then can be proved.

**Theorem 4.** *Assume that the matrix $A$ has an isotropically distributed null-space and that the number of rows of the matrix $A$ is $dm = \alpha dn$. Fix constants $c$ and $\eta$ according to (13) and arbitrarily small number $\epsilon$ and $\delta$. Let $b = \sqrt{1-\epsilon^2+c^2\epsilon^2}$, $f = c\epsilon$, and $g = \sqrt{1-\epsilon^2} - c\epsilon$. Choose $\beta < \beta_{opt}$ where $\beta_{opt} = \frac{1}{2} - O(\epsilon)$ is given by (27). For any $\mathbf{x}$ that is d-block sparse and has block sparsity $k < \beta n$, the solutions to the optimizations (1) and (3) coincide if*

$$d > \frac{H(\beta) - \ln(\frac{\eta}{\epsilon})}{\delta(\beta b + (1-\beta)f - (1-\beta)g)} \quad \text{and} \quad \alpha > 1 - \frac{1}{d} \tag{28}$$

*Proof.* Follows from the previous discussion combining (8), (22), (23), (24), (25), and (26). ∎

Before moving on to the numerical study of the performance of the algorithm (3) we should also mention that the theoretical results from [10] and [45] are related to what is often called the *strong threshold* (the interested reader can find more on the definition of the strong threshold in [10]) for sparsity. As we have said earlier, if the number of the measurements is $M = \alpha N$ then the strong threshold for sparsity is ideally $K = \frac{\alpha}{2} N$. Also, the definition of the strong threshold assumes that the reconstructing algorithm ((2), (3) or any other) succeeds for *any* sparse signal with sparsity below the strong threshold. However, since this can not be numerically verified (we simply can not generate all possible $k$ block sparse signals from $\mathbb{R}^{dn}$), a weaker notion of the threshold (called the *weak threshold*) is usually considered in numerical experiments (the interested reader can also find more on the definition of the weak threshold in [10]). The main feature of the weak threshold definition is that it allows failure in reconstruction of a certain small fraction of signals with sparsity below it. However, as expected, the ideal performance in the sense of weak threshold assumes that if the number of the measurements is $M = \alpha N$ and the sparsity is $K = \beta N$, then $\beta$ should approach $\alpha$. As the numerical experiments in the following sections hint increasing the block length $d$ leads to almost ideal performance of the reconstructing technique given in (3).

**Table 1.** The theoretical and simulation results for recovery of block-sparse signals with different block size. $\rho_S$ is the strong threshold for $\ell_1$ optimization and $\rho_W$ is the weak threshold for $\ell_1$ optimization both are found from [5, 6]. $d$ represents the block size in various simulations. The data are taken from the curves with probability of success more than %95.

|          | $\delta = 0.1$ | $\delta = 0.3$ | $\delta = 0.5$ | $\delta = 0.7$ | $\delta = 0.9$ |
|----------|----------------|----------------|----------------|----------------|----------------|
| $\rho_S$ | 0.049          | 0.070          | 0.089          | 0.111          | 0.140          |
| $\rho_W$ | 0.188          | 0.292          | 0.385          | 0.501          | 0.677          |
| $d = 1$  | 0.10           | 0.23           | 0.30           | 0.41           | 0.62           |
| $d = 4$  | 0.30           | 0.33           | 0.50           | 0.57           | 0.72           |
| $d = 8$  | 0.50           | 0.60           | 0.60           | 0.71           | 0.89           |
| $d = 16$ | 0.70           | 0.80           | 0.80           | 0.91           | 0.94           |

## 4. Numerical study of the block sparse reconstruction

In this section we recall the basics of the algorithm, show how it can efficiently be solved in polynomial time, and demonstrate its performance through numerical simulations.

In order to recover a $k$ block sparse signal $\mathbf{x}$ from the linear measurements $\mathbf{y} = A\mathbf{x}$ we consider the following optimization problem

$$\begin{aligned}
\min_{\mathbf{x}} \quad & \|\mathbf{X}_1\|_2 + \|\mathbf{X}_2\|_2 + \cdots + \|\mathbf{X}_n\|_2 \\
\text{subject to} \quad & A\mathbf{x} = \mathbf{y}
\end{aligned} \quad (29)$$

where $\mathbf{X}_i = (\mathbf{x}_{(i-1)d+1}, \mathbf{x}_{(i-1)d+2}, \ldots, \mathbf{x}_{id})$, for $i = 1, 2, \ldots, n$. Since the objective function is convex this is clearly a convex optimization problem. In principle this problem is solvable in polynomial time. Furthermore, we can transform it to a bit more convenient form in the following way

$$\begin{aligned}
\min_{\mathbf{x}, t_1, t_2, \ldots, t_n} \quad & \sum_{i=1}^{n} t_i \\
\text{subject to} \quad & \|\mathbf{X}_i\|_2^2 \leqslant t_i^2, \quad t_i \geqslant 0, \quad 1 \leqslant i \leqslant n \\
& A\mathbf{x} = \mathbf{y}
\end{aligned} \quad (30)$$

where as earlier $\mathbf{X}_i = (\mathbf{x}_{(i-1)d+1}, \mathbf{x}_{(i-1)d+2}, \ldots, \mathbf{x}_{id})$, for $i = 1, 2, \ldots, n$. Finally, it is not that difficult to see that (30) can be transformed to

$$\begin{aligned}
\min_{\mathbf{x}, t_1, t_2, \ldots, t_n} \quad & \sum_{i=1}^{n} t_i \\
\text{subject to} \quad & \begin{bmatrix} t_i I & \mathbf{X}_i^* \\ \mathbf{X}_i & t_i \end{bmatrix} \geqslant 0, \quad t_i \geqslant 0, \quad 1 \leqslant i \leqslant n \\
& A\mathbf{x} = \mathbf{y}
\end{aligned} \quad (31)$$

with $\mathbf{X}_i = (\mathbf{x}_{(i-1)d+1}, \mathbf{x}_{(i-1)d+2}, \ldots, \mathbf{x}_{id})$, for $i = 1, 2, \ldots, n$. Clearly, (31) is a semi-definite program and can be solved by a host of numerical methods in polynomial time.

To further improve the reconstruction performance we introduce an additional modification of (31). Assume that $\hat{\mathbf{X}}_i$, $1 \leqslant i \leqslant n$ is the solution of (31). Further, sort $\|\hat{\mathbf{X}}_i\|_2$ and assume that $\hat{\mathcal{K}}$ is the set of $k$ indices which correspond to the $k$ vectors $\mathbf{X}_i$ with the largest norm. Let these indices determine the positions of the nonzero

**Algorithm 1** Recovery of block-sparse signals
---
**Input:** Measured vector $\mathbf{y} \in \mathbb{R}^m$, size of blocks $d$, and measurement matrix $A$.
**Output:** Block-sparse signal $\mathbf{x} \in \mathbb{R}^n$.
1: Solve the following optimization problem

$$\min_{\mathbf{x}} \quad \|\mathbf{X}_1\|_2 + \|\mathbf{X}_2\|_2 + \cdots + \|\mathbf{X}_n\|_2$$
$$\text{subject to} \quad A\mathbf{x} = \mathbf{y}$$

   using semi-definite programming.
2: Sort $\|X_i\|_2$ for $i = 1, 2, \ldots, n$, such that $\|X_{j_1}\|_2 \geq \|X_{j_2}\|_2 \geq \cdots \geq \|X_{j_n}\|_2$.
3: The indices $j_1, j_2, \ldots, j_d$ mark the blocks of $\mathbf{x}$ that are nonzero. Set $\bar{A}$ to be the submatrix of $A$ containing columns of $A$ that are correspond to blocks $j_1, j_2, \ldots, j_d$.
4: Let $\bar{\mathbf{x}}$ represent the corresponding nonzero blocks of $\mathbf{x}$ determined by $j_1, j_2, \ldots, j_d$. Set $\bar{\mathbf{x}} = \bar{A}^{-1}\mathbf{y}$ and the rest of blocks of $\mathbf{x}$ to zero.
5: **return x**.
---

blocks. Then let $A_{\hat{\mathcal{K}}}$ be the submatrix of $A$ obtained by selecting the columns with the indices $\hat{\mathcal{K}}$ from the first $k$ rows of $A$. Also let $\mathbf{y}_{\hat{\mathcal{K}}}$ be the first $kd$ components of $\mathbf{y}$. Then we generate the nonzero part of the reconstructed signal $\hat{\mathbf{x}}$ as $\hat{\mathbf{x}}_{\hat{\mathcal{K}}} = A_{\hat{\mathcal{K}}}^{-1}\mathbf{y}_{\hat{\mathcal{K}}}$. We refer to this procedure of reconstructing the sparse signal $\mathbf{x}$ as $\ell_2/\ell_1$ algorithm and in the following subsection we show its performance.

### 4.1. Simulation results

In this section we discuss the performance of the $\ell_2/\ell_1$ algorithm. We conducted 4 numerical experiments for 4 different values of the block length $d$. In cases when $d = 1, 4$, or $8$ we set the length of the sparse vector to be $N = 800$ and in the case $d = 16$ we set $N = 1600$. For fixed values of $d$ and $N$ we then generated a random Gaussian measurement matrix $A$ for $0.1 \leq \alpha \leq 0.9$. For each of these matrices we randomly generate 100 different signals of a given sparsity $\beta$, form a measurement vector $\mathbf{y}$, and run the $\ell_2/\ell_1$ algorithm. The percentage of success (perfect recovery of the sparse signal) is shown in Figure 2 and Table 1. The case $d = 1$ corresponds to the basic $\ell_1$ relaxation. As can be seen from Figure 2 increasing the block length significantly improves the threshold for allowable sparsity.

### 5. Conclusion

In this paper we studied the efficient recovery of block sparse signals using an under-determined system of equations generated from random Gaussian matrices. Such problems arise in different applications, such as DNA microarrays, equalization of sparse communication channels, magnetoencephalography, etc. We analyzed the minimization of a mixed $\ell_2/\ell_1$ type norm, which can be reduced to solving a semi-definite program. We showed that, as the number of measurements approaches the number of unknowns, the $\ell_2/\ell_1$ algorithm can uniquely recover any block-sparse signal whose sparsity is up to half the number of measurements with overwhelming probability over the measurement matrix. This coincides with the best that can be achieved via exhaustive search. Our proof technique (which involves a certain union bound) appears to give a loose bound when the number of measurements is a fixed fraction of the number of unknowns. For future work it would be interesting to see if one could obtain "sharp" bounds on when signal recovery is possible (similar to the sharp bounds in [8]) for $\ell_2/\ell_1$ method.

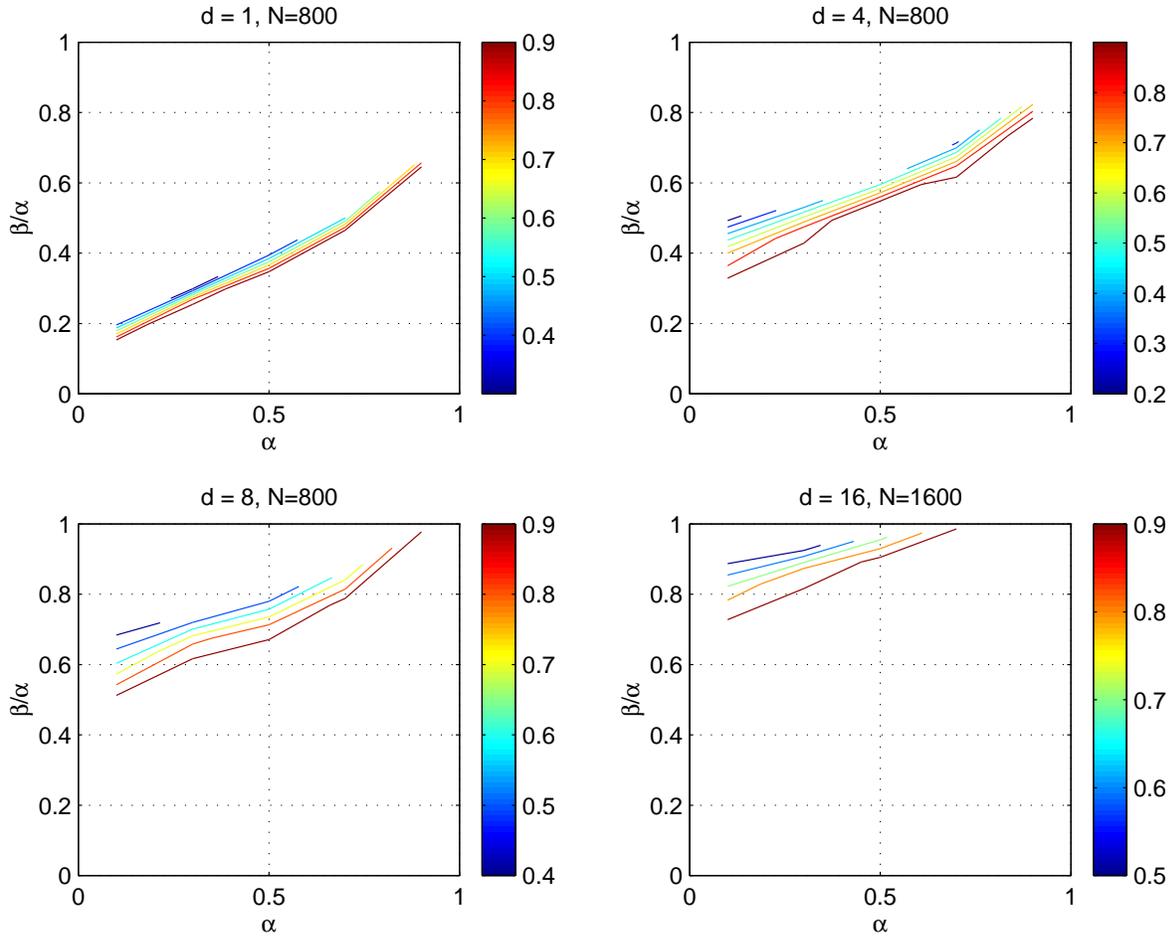

**Figure 2. Threshold for $\beta$ for a given $\alpha$ (the colors of the curves indicate the probability of success of $\ell_2/\ell_1$ algorithm calculated over $100$ independent instances of $d$-block sparse signals x and a fixed Gaussian measurement matrix $A$)**

## A. Computing $E\left[e^{\sqrt{2d-1}\delta\sqrt{2}||B_1||_2}\right]$ and $E\left[e^{\sqrt{2d-1}\delta\sqrt{2}||F_1||_2}\right]$

Now we turn to computing $Ee^{\sqrt{2d-1}\delta\sqrt{2}||B_1||_2}$ and $Ee^{\sqrt{2d-1}\delta\sqrt{2}||F_1||_2}$. Let us first consider $Ee^{\mu||B_1||_2}$. Since $B_1$ is a $d$ dimensional vector let $B_1 = [B_{1_1}, B_{1_2}, \ldots, B_{1_d}]$. As we have stated earlier $B_{1_p}, 1 \leq p \leq d$ are i.i.d. zero-mean Gaussian random variables with variance $\text{var}(B_{1_p}) = 1 - \epsilon^2 + c^2\epsilon^2 = b^2, 1 \leq p \leq d$. Then we can write

$$Ee^{\sqrt{2d-1}\delta\sqrt{2}||B_1||_2} = \frac{1}{\sqrt{2\pi}^d} \int_{-\infty}^{\infty} \cdots \int_{-\infty}^{\infty} \exp\left(\sqrt{2d-1}\delta\sqrt{2}b\sqrt{\sum_{p=1}^{d} B_{1_p}^2} - \frac{\sum_{p=1}^{d} B_{1_p}^2}{2}\right) dB_1.$$

Using the spherical coordinates it is not that difficult to show that the previous integral can be transformed to

$$\begin{aligned}
Ee^{\sqrt{2d-1}\delta\sqrt{2}||B_1||_2} &= \frac{1}{\sqrt{2\pi}^d} \frac{2\sqrt{\pi}^d}{\Gamma(\frac{d}{2})} \int_0^{\infty} r^{d-1} e^{\sqrt{2d-1}\delta\sqrt{2}br - \frac{r^2}{2}} dr \\
&= \frac{\Gamma(d) e^{\frac{(2d-1)(\delta b)^2}{2}}}{\Gamma(\frac{d}{2}) 2^{\frac{d}{2}-1}} \frac{e^{-\frac{(2d-1)(\delta b)^2}{2}}}{\Gamma(d)} \int_0^{\infty} r^{d-1} e^{\sqrt{2d-1}\delta\sqrt{2}br - \frac{r^2}{2}} dr \\
&= \frac{\Gamma(d) e^{\frac{(2d-1)(\delta b)^2}{2}}}{\Gamma(\frac{d}{2}) 2^{\frac{d}{2}-1}} U\left(\frac{2d-1}{2}, -\sqrt{2d-1}\delta\sqrt{2}b\right)
\end{aligned} \qquad (32)$$

where $U$ is *parabolic cylinder function* (see e.g., [32]). Before proceeding further we recall the asymptotic results for $U$ from [32]. Namely, from [32] we have that if $\zeta \gg 0$ and $t \geq 0$

$$U\left(\frac{\zeta^2}{2}, -\zeta t\sqrt{2}\right) \approx \frac{h(\zeta) e^{\zeta^2 \tilde{\rho}} \sqrt{2\pi}}{\Gamma(\frac{\zeta^2+1}{2})(t^2+1)^{\frac{1}{4}}} \qquad (33)$$

where

$$h(\zeta) = 2^{-\frac{\zeta^2}{4} - \frac{1}{4}} e^{-\frac{\zeta^2}{4}} \zeta^{\frac{\zeta^2}{2} - \frac{1}{2}}, \quad \tilde{\rho} = \frac{1}{2}(t\sqrt{1+t^2} + \ln(t + \sqrt{1+t^2})). \qquad (34)$$

¿From (33) and (34) we have

$$U\left(\frac{2d-1}{2}, -\sqrt{2d-1}\delta b\sqrt{2}\right) \approx \frac{1}{\Gamma(d)} \left((2e)^{-\frac{2d-1}{4}} \sqrt{2d-1}^{\frac{2d-1}{2} - \frac{1}{2}} 2^{-\frac{1}{4}}\right) \frac{e^{\frac{2d-1}{2}(\frac{1}{2}(\delta b\sqrt{1+(\delta b)^2} + \ln(\delta b + \sqrt{1+(\delta b)^2})))}}{(1+(\delta b)^2)^{\frac{1}{4}}}. \qquad (35)$$

Connecting (32) and (35) we finally obtain for $d \gg 0$ and $\delta \ll 1$ ($\delta$ is a constant independent of $d$)

$$Ee^{\sqrt{2d-1}\delta\sqrt{2}||B_1||_2} \approx \frac{e^{\frac{2d-1}{2}(\delta b)^2}}{\Gamma(\frac{d}{2}) 2^{\frac{d}{2}-1}} \left((2e)^{-\frac{2d-1}{4}} \sqrt{2d-1}^{\frac{2d-1}{2} - \frac{1}{2}} 2^{-\frac{1}{4}}\right) \frac{e^{\frac{2d-1}{2}(\frac{1}{2}(\delta b\sqrt{1+(\delta b)^2} + \ln(\delta b + \sqrt{1+(\delta b)^2})))}}{(1+(\delta b)^2)^{\frac{1}{4}}}. \qquad (36)$$

Using the facts that $\delta b \ll 1$ and $\Gamma(\frac{d}{2}) \approx (\frac{d}{2e})^{\frac{d}{2}}$ when $d$ is large, (36) can be rewritten as

$$Ee^{\sqrt{2d-1}\delta\sqrt{2}||B_1||_2} \approx e^{\frac{2d-1}{2}(\delta b)^2} e^{\frac{2d-1}{2}(\frac{1}{2}(\delta b\sqrt{1+(\delta b)^2}+\ln(\delta b+\sqrt{1+(\delta b)^2})))}.$$

Since $\delta b \ll 1$ it further follows

$$\begin{aligned}
Ee^{\sqrt{2d-1}\delta\sqrt{2}||B_1||_2} &\approx e^{\frac{2d-1}{2}((\delta b)^2+\frac{1}{2}(\delta b\sqrt{1+(\delta b)^2}+\ln(\delta b+\sqrt{1+(\delta b)^2})))} \\
&\approx e^{\frac{2d-1}{2}((\delta b)^2+\frac{1}{2}(\delta b(1+\frac{(\delta b)^2}{2})+\ln(\delta b+(1+\frac{(\delta b)^2}{2}))))} \\
&\approx e^{\frac{2d-1}{2}((\delta b)^2+\frac{1}{2}(\delta b(1+\frac{(\delta b)^2}{2})+\delta b))} \\
&\approx e^{\frac{2d-1}{2}((\delta b)^2+\delta b)} \\
&\approx e^{d((\delta b)^2+\delta b)}. \tag{37}
\end{aligned}$$

To compute $Ee^{\mu ||F_1||_2}$ we first note that $F_1$ is a $d$ dimensional vector. Let $F_1 = [F_{1_1}, F_{1_2}, \ldots, F_{1_d}]$. As we have stated earlier $F_{1_p}, 1 \leq p \leq d$ are i.i.d. zero-mean Gaussian random variables with variance $\text{var}(F_{1_p}) = c^2\epsilon^2 = f^2, 1 \leq p \leq d$. Then the rest of the derivation for computing $Ee^{\sqrt{2d-1}\delta\sqrt{2}||F_1||_2}$ follows directly as in the case of $Ee^{\sqrt{2d-1}\delta\sqrt{2}||B_1||_2}$. Hence we can write similarly to (37)

$$Ee^{\sqrt{2d-1}\delta\sqrt{2}||F_1||_2} \approx e^{d((\delta f)^2+\delta f)}. \tag{38}$$

## B. Computing $E\left[e^{-\sqrt{2d-1}\delta\sqrt{2}||G_1||_2}\right]$

Now we turn to computing $Ee^{-\sqrt{2d-1}\delta\sqrt{2}||G_1||_2}$. Since $G_1$ is a $d$ dimensional vector let $G_1 = [G_{1_1}, G_{1_2}, \ldots, G_{1_d}]$. As we have stated earlier $G_{1_p}, 1 \leq p \leq d$ are i.i.d. zero-mean Gaussian random variables with variance $\text{var}(G_{1_p}) = (\sqrt{1-\epsilon^2} - c\epsilon)^2 = g^2, 1 \leq p \leq d$. Then we can write

$$Ee^{-\sqrt{2d-1}\delta\sqrt{2}||G_1||_2} = \frac{1}{\sqrt{2\pi}^d} \int_{-\infty}^{\infty} \cdots \int_{-\infty}^{\infty} \exp\left(-\sqrt{2d-1}\delta\sqrt{2}g\sqrt{\sum_{p=1}^{d} G_{1_p}^2} - \frac{\sum_{p=1}^{d} G_{1_p}^2}{2}\right) dG_1.$$

Similarly as in the previous subsection using the spherical coordinates it is not that difficult to show that the previous integral can be transformed to

$$\begin{aligned}
Ee^{-\sqrt{2d-1}\delta\sqrt{2}||G_1||_2} &= \frac{1}{\sqrt{2\pi}^d}\frac{2\sqrt{\pi}^d}{\Gamma(\frac{d}{2})} \int_0^{\infty} r^{d-1} e^{-\sqrt{2d-1}\delta\sqrt{2}gr - \frac{r^2}{2}} dr \\
&= \frac{\Gamma(d)e^{\frac{(2d-1)(\delta g)^2}{2}}}{\Gamma(\frac{d}{2})2^{\frac{d}{2}-1}} \frac{e^{-\frac{(2d-1)(\delta g)^2}{2}}}{\Gamma(d)} \int_0^{\infty} r^{d-1} e^{-\sqrt{2d-1}\delta\sqrt{2}gr - \frac{r^2}{2}} dr \\
&= \frac{\Gamma(d)e^{\frac{(2d-1)(\delta g)^2}{2}}}{\Gamma(\frac{d}{2})2^{\frac{d}{2}-1}} U\left(\frac{2d-1}{2}, \sqrt{2d-1}\delta\sqrt{2}g\right) \tag{39}
\end{aligned}$$

where as earlier $U$ is *parabolic cylinder function*. Before proceeding further we again recall another set of the asymptotic results for $U$ from [32]. Namely, from [32] we have that if $\zeta \gg 0$ and $t \geq 0$

$$U\left(\frac{\zeta^2}{2}, \zeta t\sqrt{2}\right) \approx \frac{\tilde{h}(\zeta)e^{-\zeta^2\tilde{\rho}}}{(t^2+1)^{\frac{1}{4}}} \tag{40}$$

where
$$\tilde{h}(\zeta) = 2^{\frac{\zeta^2}{4}-\frac{1}{4}} e^{\frac{\zeta^2}{4}} \zeta^{-\frac{\zeta^2}{2}-\frac{1}{2}}, \quad \tilde{\rho} = \frac{1}{2}(t\sqrt{1+t^2} + \ln(t+\sqrt{1+t^2})). \tag{41}$$

¿From (40) and (41) we have

$$U(\frac{2d-1}{2}, \sqrt{2d-1}\delta g\sqrt{2}) \approx \left((2e)^{\frac{2d-1}{4}}\sqrt{2d-1}^{-\frac{2d-1}{2}-\frac{1}{2}} 2^{-\frac{1}{4}}\right) \frac{e^{-\frac{2d-1}{2}(\frac{1}{2}(\delta g\sqrt{1+(\delta g)^2}+\ln(\delta g+\sqrt{1+(\delta g)^2})))}}{(1+(\delta g)^2)^{\frac{1}{4}}}. \tag{42}$$

Connecting (39) and (42) we finally obtain for $d \gg 0$ and $\delta \ll 1$ (as earlier $\delta$ is a constant independent of $d$)

$$Ee^{-\sqrt{2d-1}\delta\sqrt{2}\|G_1\|_2} \approx \frac{\Gamma(d) e^{\frac{2d-1}{2}(\delta g)^2}}{\Gamma(\frac{d}{2}) 2^{\frac{d}{2}-1}} \left((2e)^{\frac{2d-1}{4}}\sqrt{2d-1}^{-\frac{2d-1}{2}-\frac{1}{2}} 2^{-\frac{1}{4}}\right) \frac{e^{-\frac{2d-1}{2}(\frac{1}{2}(\delta g\sqrt{1+(\delta g)^2}+\ln(\delta g+\sqrt{1+(\delta g)^2})))}}{(1+(\delta g)^2)^{\frac{1}{4}}}. \tag{43}$$

Using the facts that $\Gamma(\frac{d}{2}) \approx (\frac{d}{2e})^{\frac{d}{2}}$ and $\Gamma(d) \approx (\frac{d}{e})^d$ when $d$ is large, (43) can be rewritten as

$$Ee^{-\sqrt{2d-1}\delta\sqrt{2}\|G_1\|_2} \approx e^{-\frac{2d-1}{2}(\delta g)^2} e^{\frac{2d-1}{2}(\frac{1}{2}(\delta g\sqrt{1+(\delta g)^2}+\ln(\delta g+\sqrt{1+(\delta g)^2})))}.$$

Since $\delta g \ll 1$ it further follows

$$\begin{aligned}
Ee^{-\sqrt{2d-1}\delta\sqrt{2}\|G_1\|_2} &\approx e^{\frac{2d-1}{2}((\delta g)^2 - \frac{1}{2}(\delta g\sqrt{1+(\delta g)^2}+\ln(\delta g+\sqrt{1+(\delta g)^2})))} \\
&\approx e^{\frac{2d-1}{2}((\delta g)^2 - \frac{1}{2}(\delta g(1+\frac{(\delta g)^2}{2})+\ln(\delta g+(1+\frac{(\delta g)^2}{2}))))} \\
&\approx e^{\frac{2d-1}{2}((\delta g)^2 - \frac{1}{2}(\delta g(1+\frac{(\delta g)^2}{2})+\delta g))} \\
&\approx e^{\frac{2d-1}{2}((\delta g)^2 - \delta g)} \\
&\approx e^{d((\delta g)^2 - \delta g)}. \tag{44}
\end{aligned}$$